\documentclass[hidelinks,12pt,a4paper]{article}

\usepackage[left=2.5cm,right=2.5cm,
top=2.5cm,bottom=2.5cm
]{geometry}
\usepackage[super,sort&compress,comma,numbers]{natbib}
\usepackage{amsmath}
\usepackage{newtxmath}
\usepackage{newtxtext}
\usepackage{bm}
\usepackage{booktabs}
\usepackage{adjustbox} 
\usepackage[ruled,vlined]{algorithm2e} 
\usepackage{etoolbox} 
\BeforeBeginEnvironment{algorithm}{\vspace{1em}}
\AfterEndEnvironment{algorithm}{\vspace{1em}}

\usepackage{graphicx}
\usepackage{float}
\usepackage{siunitx}
\usepackage[dvipsnames]{xcolor}
\usepackage[none]{hyphenat}
\usepackage{indentfirst}
\usepackage[format=plain, justification=justified, singlelinecheck=false, font=small, labelfont=bf, labelsep=period]{caption}
\usepackage{hyperref} 
\usepackage{xr-hyper}
\usepackage{setspace}
\graphicspath{{figures/}}

\definecolor{RevisionColour}{named}{RoyalBlue}
\definecolor{RevisionColour}{named}{black}

\begin{document}

{\centering{\LARGE\textbf{Finite elements and moving asymptotes accelerate quantum optimal control --- FEMMA}\par}}\bigbreak

\begin{center}
Mengjia He\textit{$^{1}$}, Yongbo Deng\textit{$^{1}$}, Burkhard Luy\textit{$^{2,3}$} and Jan G. Korvink\textit{$^{1,*}$}
\bigbreak

\textit{$^{1}$Institute of Microstructure Technology, Karlsruhe Institute of Technology, Eggenstein-Leopoldshafen, Germany.}

\textit{$^{2}$Institute for Biological Interfaces 4 - Magnetic Resonance, Karlsruhe Institute of Technology, Eggenstein-Leopoldshafen, Germany.}

\textit{$^{3}$Institute of Organic Chemistry, Karlsruhe Institute of Technology, Karlsruhe, Germany.}
\bigbreak

\text{$^{*}$ Corresponding author: jan.korvink@kit.edu}
\bigbreak
\end{center}

{\centering \textbf{Abstract}\bigbreak\noindent
Quantum optimal control is central to designing spin manipulation pulses. \textcolor{RevisionColour}{Gradient-based pulse optimization can be facilitated by either accelerating gradient evaluation, or enhancing the convergence rate.} In this work, we accelerated single-spin optimal control by combining the finite element method with the method of moving asymptotes. By treating discretized time as spatial coordinates, the Liouville–von Neumann equation was reformulated as a linear system, \textcolor{RevisionColour}{efficiently yielding a joint solution of the spin trajectory and control gradient. The method of moving asymptotes, relying on the ensemble fidelities and gradients, achieves rapid convergence for a target fidelity of 0.995.}}

\section{Introduction}
\label{sec:intro}
In magnetic resonance spectroscopy (MRS) and imaging (MRI), shaped radio frequency (RF) pulses are widely employed to drive the state of a spin system toward a desired target. The design of such pulses is typically framed as an optimal control problem~\cite{kochQuantum2022a}. 

\textcolor{RevisionColour}{Among gradient-based optimization techniques, the gradient ascent pulse engineering (GRAPE) method~\cite{khanejaOptimal2005} has become one of the most widely used and efficient methods. The original GRAPE method discretizes the control time into piecewise-constant intervals and computes the spin trajectory via a bi-directional propagation, where gradients are obtained using a first-order finite-difference approximation under sufficiently fine time slicing.}

\textcolor{RevisionColour}{A variety of extensions to GRAPE have since been developed, targeting more accurate derivative computation, advanced optimization strategies, and broader applicability to general quantum systems. The derivatives of the sub-propagator to control variables have been derived by differentiating the Taylor series~\cite{kuprovDerivatives2009}. A higher-order finite difference method was introduced to improve gradient accuracy~\cite{defouquieres2011Second}. The auxiliary matrix formalism (AUXMAT) was introduced to compute exact gradients and Hessians, leading to improved convergence~\cite{goodwin2015Auxiliary}. Automatic differentiation, implemented on graphics processing units, was introduced to accelerate gradient computation~\cite{leung2017speedup,abdelhafez2019gradient}. Beyond the gradient-descent update of the original formulation, several optimization algorithms have been incorporated into GRAPE frameworks, including Quasi-Newton approaches~\cite{defouquieres2011Second,machnesComparing2011,jensenApproximate2021}, such as the Broyden–Fletcher–Goldfarb–Shanno (BFGS) algorithm and its limited-memory version (L-BFGS), and Newton–Raphson methods~\cite{goodwin2016Modified} that provide quadratic convergence. The accelerated Newton–Raphson variant was developed to reduce the Hessian evaluation to $O(N)$ complexity, as demonstrated for a single-spin system~\cite {Goodwin:2022uwc}. Extension algorithms for open quantum systems demonstrated high-fidelity control when handling decoherence effects~\cite{abdelhafez2019gradient,Chen2025Robust}.}

As modern magnetic resonance systems operate at increasingly higher frequencies, various instrumental limitations have become more prominent, including RF power constraints \cite{kobzarExploring2004,kobzarExploring2008,josephOptimal2023} and hardware-induced distortions \cite{tosnerOvercoming2018,rasulov2023Simulation,Lowe:2024fbg,rasulov2025Instrumental}. Optimization problems also grow more complex when modeling multi-qubit systems exhibiting entanglement \cite{Goodwin:2022thz}, heteronuclear spin systems with $J$-coupling \cite{ehniBEBEtr2013}, or crystalline orientation distributions in solid-state MRS \cite{tosnerMaximizing2021}. In parallel transmit MRI, pulse design under safety constraints can involve thousands of control variables and hundreds of spatial voxels \cite{zhu2004Parallel,vinding2017Local,vinding2019Ultrafast}. To address crosstalk in parallel MRS, compensation strategies incorporate an ensemble of $B_0$ distortions caused by gradient coils \cite{he2025Coherence}, along with multiple cooperative pulses tailored for parallel excitation \cite{he2024Digital}.

These scenarios often require ensemble-based optimization, where system parameters vary across ensemble members. In such cases, the gradient or Hessian must be computed for each ensemble element, and control updates are determined from the ensemble-averaged derivatives, as in the L-BFGS and Newton methods. This significantly increases the computational cost, potentially requiring several hours or even days of high-performance computing time.

In this work, we address the optimization of single-spin magnetic resonance pulses by combining the finite element method (FEM)~\cite{hughesFinite2012,StrangFix1973} with the method of moving asymptotes (MMA) \cite{svanberg1987Method}. \textcolor{RevisionColour}{Finite elements with linear, quadratic, and cubic Hermite basis functions provide an efficient and scalable solution for computing single-spin optimal-control gradients.} By formulating the ensemble-averaged fidelity maximization as a constrained optimization, MMA converges more rapidly than L-BFGS and yields the lowest overall runtime among L-BFGS and Newton–Raphson methods.

\section{Methodology}
The Liouville-von Neumann (LvN) equation masters the evolution of a general spin system. The equation in Liouville space is given by
\begin{equation}
\label{LvN}
\begin{aligned}
\frac{d}{dt}\bm{\rho}(t) + \mathrm{i}\mathbf{L}\bm{\rho}(t) &= 0, \\
\bm{\rho}(0) &= \bm{\rho}_0,
\end{aligned}
\end{equation}
where $\bm{\rho}_0 $ is the initial state. While ignoring the relaxation effect, the Liouvillian $\mathbf{L} $ can be decomposed into the internal part and the control part,
\begin{align}
\label{eq-hamiltonain}
\mathbf{L}(t) = \mathbf{L}_{\rm{int}}(t)+\sum\limits_{m}x_m(t)\mathbf{L}_m,
\end{align}
where $\mathbf{L}_{\mathrm{int}}$ could contain Zeeman interaction with the magnetic field and spin-spin couplings. The $\mathbf{L}_m$ represents a control operator and $x_m(t)$ is a time-dependent coefficient. A control sequence $\bm{x}(t)$ is applied to steer the spin system from an initial state $\bm{\rho_0}$ to a target state $\bm{\sigma}$ in a specified time duration $T$. A measurement of the control efficiency is the overlap between the target state and the actual final state $\bm{\rho_T}$, i.e., 
\begin{align}
\label{eqs-fidelity}
\eta= \left<{\bm{\sigma|\rho_T}}\right>,
\end{align}
which satisifies $-1 \leq \eta \leq 1$. We consider an ensemble spin system with $N_{\text{ens}}$ members, for which the control amplitudes are limited, so that the optimal control problem can be defined as:
\begin{equation}
\label{eq-optimalModel}
\left\{
\begin{aligned}
& \text{Find } \bm{x}(t), \, t \in [0,T],\\
& \text{to maximize } f = \sum\limits_{k=1}^{N_{\text{ens}}} \eta_k, \\
& \text{constrained by } |x_m(t)| \leq x_m^{\text{max}}.
\end{aligned}
\right.
\end{equation}

\subsection{FEM solution of the Liouville-von Neumann equation}
\label{FEM}
\begin{figure}[H]
    \centering
    \includegraphics[width=0.8\textwidth]{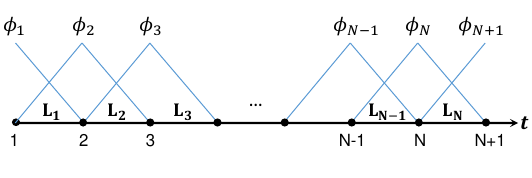}
    \caption{View of the one-dimensional linear Lagrange global shape functions. The time interval between two nodes is uniformly set to $\Delta t$. The labels $\mathbf{L}_1, \mathbf{L}_2, \ldots, \mathbf{L}_N$ represent the discretized Liouvillian for a piecewise-constant waveform.}
    \label{fig-femMesh}
\end{figure}

In the context of magnetic resonance, the FEM has previously been applied to solve the stochastic Liouville equation for chemically induced spin polarization problems \cite{zientara1979Variational,zientara1979Variationala} and for electron spin resonance spectral simulations \cite{stillman1979Variational}. In Eq.~\ref{LvN}, the spatial variables are absent relative to the stochastic Liouville equation. Within the Hamilton principle, the integration domain, time interval $[0, T]$, is discretized into $N$ elements with $N+1$ nodes. The control variables are represented as a piecewise-constant waveform $\mathbf{x}$, rendering the Liouvillian $\mathbf{L}$ constant within each element. The solution of Eq.~\ref{LvN} is approximated as a linear combination of shape functions,
\begin{align}
\label{rhobyLinearCombine}
\bm{\rho}(t) = \sum_{j=1}^{N+1} \bm{\alpha}_j \phi_j(t),
\end{align}
where $\bm{\alpha}_j$ has the same dimension as the spin state vector $\bm{\rho}(t)$, and $\phi_j(t)$ represents the $j$-th shape function. As shown in Fig.~\ref{fig-femMesh}, by using the linear elements, the nodal shape functions globally defined at node $j$ are expressed as
\begin{align}
\label{globalShapeFunction}
\phi_j^n(t) =
\begin{cases} 
(t - t_{j-1})/\Delta t, & t_{j-1} \leq t \leq t_j, \\
(t_{j+1} - t)/\Delta t, & t_j \leq t \leq t_{j+1}, \\
0, & \text{otherwise}.
\end{cases}
\end{align}

Note that it's also possible to define the two linear shape functions locally at element $j$ as 
\begin{align}
\color{RevisionColour}
\label{localShapeFunction}
\phi^{e1}_j(t)=(t - t_{j})/\Delta t, \quad
\phi^{e2}_j(t)=(t_j+\Delta t - t)/\Delta t,  \quad t_{j} \leq t \leq t_j+\Delta t.
\end{align}
In the following derivation, globally defined shape functions are used, and the subscript $n$ is omitted.

The Galerkin's method~\cite{Galerkin1915} demonstrated that the integral of the weighted residual equals zero, i.e.,
\begin{align}
\label{weakForm}
\int_{0}^{T} \left(\frac{d \bm{\rho}}{dt} + \text{i}\mathbf{L}\bm{\rho} \right) \hat{\omega} \, dt = 0.
\end{align}
where $\bm{\rho}$ is the approximated solution expressed by Eq.~\ref{rhobyLinearCombine}, $\hat{\omega}$ is the chosen weighting function. Selecting each shape function as the weighting function, i.e., $\hat{\omega} = \phi_j$, and substituting Eq.~\ref{rhobyLinearCombine} into Eq.~\ref{weakForm} gives the following linear equations,
\begin{align}
\label{rhoLinearEqs}
\sum_{j=1}^{N+1} \bm{\alpha}_j \int_{0}^{T} \left( \frac{d \phi_j}{dt} + \text{i}\mathbf{L} \phi_j \right) \phi_i \, dt = 0, \quad i = 1, \dots, N+1.
\end{align}
In matrix form, this can be written compactly as:
\begin{align}
\label{rhoFEM}
\mathbf{K} \cdot \bm{\alpha} = \mathbf{f},
\end{align}
where $\bm{\alpha}$ is the vectorized $\bm{\alpha}_j$ ($j=1,\,2,...,\,N+1$), $\mathbf{f} = 0$ is a forcing function, often called the load vector, and $\mathbf{K}$ is an impedance function, often called the stiffness matrix. The terms load and stiffness appeared first in the finite element literature because of the application to computational mechanics. An element of $\mathbf{K}$ is  given by:
\begin{align}
\label{stiffElement}
K_{ij} = \int_{0}^{T} \left( \frac{d \phi_j}{dt} + \text{i}\mathbf{L} \phi_j \right) \phi_i \, dt = \sum_{n=1}^{N} \int_{t_n}^{t_{n+1}} \left( \frac{d \phi_j}{dt} + \text{i}\mathbf{L} \phi_j \right) \phi_i \, dt.
\end{align}
Considering that $\phi_j$ is nonzero only within the elements connected to node $j$, the global stiffness matrix $\mathbf{K}$ can be assembled from the element stiffness matrices:
\begin{align}
\label{rhoStiff}
\mathbf{K} =
\begin{bmatrix}
K_{11}^{e,1} & K_{12}^{e,1} &  &  &  &  \\
K_{21}^{e,1} & K_{22}^{e,1}+K_{11}^{e,2} & K_{12}^{e,2} &  &  & \\
 & K_{21}^{e,2} & K_{22}^{e,2} + K_{11}^{e,3} & K_{12}^{e,3} & &  \\
 &  & K_{21}^{e,3} & \ddots  &  &  \\
&  &  &  & K_{22}^{e,N-1}+K_{11}^{e,N} & K_{12}^{e,N}  \\
 &  &  &  & K_{21}^{e,N}  & K_{22}^{e,N}
\end{bmatrix},
\end{align}
which reveals its banded structure, while the stiffness matrix of the $j$-th element is given by
\begin{align}
\label{rhoEleStiff_integral}
K^{e,j} =
\begin{bmatrix}
\int_{t_j}^{t_{j+1}} \left( \frac{d \phi_j}{dt} + \mathrm{i}\mathbf{L} \phi_j \right) \phi_j \, dt & \int_{t_j}^{t_{j+1}} \left( \frac{d \phi_{j+1}}{dt} + \mathrm{i}\mathbf{L} \phi_{j+1} \right) \phi_j \, dt \\
\int_{t_j}^{t_{j+1}} \left( \frac{d \phi_j}{dt} + \mathrm{i}\mathbf{L} \phi_j \right) \phi_{j+1} \, dt & \int_{t_j}^{t_{j+1}} \left( \frac{d \phi_{j+1}}{dt} + \mathrm{i}\mathbf{L} \phi_{j+1} \right) \phi_{j+1} \, dt
\end{bmatrix}.
\end{align}
Substituting the expressions for the linear shape functions yields:
\begin{align}
\label{rhoEleStiff}
K^{e,j} =
\begin{bmatrix}
-\frac{\mathbf{E}}{2} + \frac{\mathrm{i}\mathbf{L}}{3} \Delta t & \frac{\mathbf{E}}{2} + \frac{\mathrm{i}\mathbf{L}}{6} \Delta t \\
-\frac{\mathbf{E}}{2} + \frac{\mathrm{i}\mathbf{L}}{6} \Delta t & \frac{\mathbf{E}}{2} + \frac{\mathrm{i}\mathbf{L}}{3} \Delta t
\end{bmatrix}.
\end{align}
Equation~\ref{rhoEleStiff} defines the element stiffness matrix derived from a general Liouvillian. For a single-spin system, the nodal spin vector is $\bm{\rho} \in \mathbb{C}^{4 \times 1}$, the nodal Liouvillian is $\mathbf{L} \in \mathbb{C}^{4 \times 4}$, and $\mathbf{E}$ denotes the identity matrix. Each element stiffness matrix $K^{e,j} \in \mathbb{C}^{8 \times 8}$ couples two nodes, while the corresponding global stiffness matrix is $\mathbf{K} \in \mathbb{C}^{(4N+4) \times (4N+4)}$. The spin trajectory is obtained through four steps, presented in Algorithm \ref{solveFEMrho}.

\begin{algorithm}[H]
\label{solveFEMrho}
\caption{Solve the linear system.}
\SetAlgoLined
\KwIn{Initial state $\bm{\rho}_0$, stiffness matrix $\mathbf{K}$ and load vector $\mathbf{f}$}
\KwOut{Solution vector $\bm{\alpha}$}
\BlankLine
Initialize $\bm{\alpha}_{1:4} \gets \bm{\rho}_0$\;
Update load vector: $\mathbf{f} \gets \mathbf{f} - \mathbf{K}_{[:,\,1{:}4]} \bm{\rho}_0$\;
Modify stiffness matrix: 
$\mathbf{K} \gets
\begin{bmatrix}
\mathbb{E}_4 & \\
 & \mathbf{K}_{[5{:}4N+4,\,5{:}4N+4]}
\end{bmatrix}$\;
Solve linear system: $\mathbf{K} \bm{\alpha} = \mathbf{f}$\;
\end{algorithm}

\textcolor{RevisionColour}{The above formulation is based on linear shape functions, but higher-order elements can be used to improve accuracy. For example, quadratic elements provide a finer discretization of the time axis but double the size of the stiffness matrix compared with linear elements. When a piecewise-linear waveform is employed to mitigate instrumental distortions~\cite{rasulov2023Simulation}, the Hamiltonian must remain continuous across time intervals. Hermite shape functions enforce $C^1$ continuity between elements and are therefore well suited to this purpose. The stiffness matrices corresponding to quadratic elements and Hermite elements are given in the supplementary material, Sections S1 and S2, respectively.}

\subsection{Adjoint analysis}
For compatibility with the FEM solution, the fidelity function in Eq.~\ref{eqs-fidelity} is rewritten as
\begin{align}
\label{eq-objective}
\eta=\mathbf{\sigma^\dagger} \cdot \rho_T = \left[ \begin{array}{cc} \mathbf{0} & \mathbf{\sigma^\dagger} \end{array} \right] \cdot \bm{\alpha},
\end{align}
where $\bm{\rho}_T = \bm{\alpha}_{4N+1:4N+4}$ denotes the final state. The gradient of the objective with respect to the control variables $\mathbf{x}$ is
\begin{align}
\label{eq-gradient}
\frac{d \eta}{d \mathbf{x}}=\frac{d \eta}{d \bm{\alpha}}\frac{d \bm{\alpha}}{d \mathbf{x}}=\frac{d \eta}{d \bm{\alpha}}\mathbf{K}^{-1}\left(\frac{d \mathbf{f}}{d \mathbf{x}}-\frac{d \mathbf{K}}{d \mathbf{x}}\bm{\alpha} \right).
\end{align}
On the right-hand side, evaluating the third term ${d\mathbf{f}}/{d\mathbf{x}} -(d\mathbf{K}/d\mathbf{x}) \bm{\alpha}$ requires $N$ matrix–vector multiplications. Since the finite elements are aligned with the discrete waveform, over which the Hamiltonian is constant, this term can be computed locally on each element, using the trajectory associated with the corresponding nodes. For instance, when $j=1$,
\begin{align}
\label{eq-dKdx1}
\frac{\partial\mathbf{f}}{\partial x_1}-\frac{\partial\mathbf{K}}{\partial x_1}\bm{\alpha}=\begin{bmatrix}
-\frac{\partial}{\partial x_1}\begin{bmatrix} K^{e,j}_{11} & \mathbf{0} \\ K^{e,1}_{21} & K^{e,1}_{22} \end{bmatrix} &  \\
& \mathbf{0} \\
\end{bmatrix}\begin{bmatrix}\alpha_{1:8}\\\mathbf{0} \end{bmatrix},
\end{align}
when $j \geq 2$,
\begin{align}
\label{eq-dKdxj}
\frac{\partial\mathbf{f}}{\partial x_j}-\frac{\partial\mathbf{K}}{\partial x_j}\bm{\alpha}=\begin{bmatrix}
\mathbf{0} & & \\
& -\frac{\partial K^{e,j}}{\partial x_j} & \\
& & \mathbf{0}
\end{bmatrix}\begin{bmatrix}\mathbf{0}\\\alpha_{4j-3:4j+4} \\ \mathbf{0} \end{bmatrix}.
\end{align}
The ${\partial K^{e,j}}/{\partial x_j}$ can be obtained by differentiating $K^{e,j}$ with respect to $x_j$ in Eq.~\ref{rhoEleStiff}. Subsequently, multiplying $\mathbf{K}^{-1}$ by $N$ vectors can be efficiently performed using the adjoint method \cite{cao2003Adjoint}, where an adjoint equation is defined as
\begin{align}
\label{eq-adjointEquation}
\mathbf{K}^{\mathsf{T}} \bm{\lambda}=\left(\frac{d \eta}{d \bm{\alpha}} \right)^{\mathsf{T}},
\end{align}
where $\bm{\lambda} \in \mathbb{C}^{(4N+4) \times 1}$ is the adjoint vector. After obtaining $\bm{\lambda}$, the gradient is computed as
\begin{align}
\label{eq-gradientByAdjoint}
\frac{d \eta}{d \mathbf{x}}=\bm{\lambda}^{\mathsf{T}} \left(\frac{d \mathbf{f}}{d \mathbf{x}}-\frac{d \mathbf{K}}{d \mathbf{x}}\bm{\alpha} \right).
\end{align}
With the adjoint method, the cost of gradient computation becomes comparable to that of solving the spin trajectory. 

\subsection{Regularization}
To ensure smooth pulse shapes, we apply the Helmholtz filter \cite{lazarov2011Filters}, defined as
\begin{align}
\label{helmFilter}
x^s(t) = R^2 \frac{d^2 x^s(t)}{dt^2} + x^c(t),
\end{align}
where $R$ denotes the filter radius. The term $x^c(t)$ corresponds to the original control variables, while $x^s(t)$ denotes the filtered (smoothed) variables. By approximating the solution with linear shape functions, Eq.~\ref{helmFilter} is solved through a separate linear system:
\begin{align}
\label{helmFEM}
\mathbf{K}_h \mathbf{x}^s = \mathbf{f}_h,
\end{align}
where the stiffness matrix elements are
\begin{align}
\color{RevisionColour}
\label{HelmstiffElement}
K_{ij} = \int_0^T  R^2 \frac{d \phi_j}{dt} \frac{d \phi_i}{dt} + \phi_j \phi_i dt,
\end{align}
and the load vector elements are
\begin{align}
\label{helmloadVector_integral}
f_i = \int_0^T x^c(t) \phi_i(t) dt.
\end{align}
The element stiffness matrix over interval $[t_j, t_{j+1}]$ is
\begin{align}
\color{RevisionColour}
\label{helmstiffEle_integral}
K^{e,j} =
\begin{bmatrix}
\int_{t_j}^{t_{j+1}}R^2\frac{d \phi_j}{dt}\frac{d \phi_j}{dt}+\phi_j\phi_j \, dt & \int_{t_j}^{t_{j+1}}R^2\frac{d \phi_{j+1}}{dt}\frac{d \phi_j}{dt}+\phi_{j+1}\phi_j \, dt \\
\int_{t_j}^{t_{j+1}}R^2\frac{d \phi_j}{dt}\frac{d \phi_{j+1}}{dt}+\phi_j\phi_{j+1} \, dt & \int_{t_j}^{t_{j+1}}R^2\frac{d \phi_{j+1}}{dt}\frac{d \phi_{j+1}}{dt}+\phi_{j+1}\phi_{j+1} \, dt
\end{bmatrix}.
\end{align}
By substituting linear shape functions, the equation simplifies to
\begin{align}
\color{RevisionColour}
\label{helmEleStiff}
K^{e,j} =
\begin{bmatrix}
\frac{\Delta t}{3} + \frac{R^2}{\Delta t} & \frac{\Delta t}{6} - \frac{R^2}{\Delta t} \\
\frac{\Delta t}{6} - \frac{R^2}{\Delta t} & \frac{\Delta t}{3} + \frac{R^2}{\Delta t}
\end{bmatrix}.
\end{align}
Since the control variables $\mathbf{x}^c$ are piecewise constant, the load vector is computed as
\begin{align}
\label{helmLoadVector}
f_i = \begin{cases}
\frac{\Delta t}{2} x^c_1, & i=1, \\
\frac{\Delta t}{2} (x^c_{i-1} + x^c_i), & 2 \leq i \leq N, \\
\frac{\Delta t}{2} x^c_N, & i = N+1.
\end{cases}
\end{align}
The Jacobian matrix that relates the smoothed variables to the control variables is
\begin{align}
\label{helmGradient}
\frac{d \mathbf{x}^s}{d \mathbf{x}^c} = \mathbf{K}_h^{-1} \left( \frac{\partial \mathbf{f}_h}{\partial x^c_1}, \frac{\partial \mathbf{f}_h}{\partial x^c_2}, \ldots, \frac{\partial \mathbf{f}_h}{\partial x^c_N} \right),
\end{align}
which yields
\begin{align}
\label{helmGradient2}
\frac{d \mathbf{x}^s}{d \mathbf{x}^c} = \frac{\Delta t}{2} [\mathbf{K}_h^{-1}]_{[1:N, 1:N+1]} {\begin{bmatrix}
1 & & & & \\
1 & 1 & & & \\
& 1 & \ddots & & \\
& & \ddots & 1 & \\
& & & 1 & 1 \\
& & & & 1
\end{bmatrix}.}_{(N+1) \times N}
\end{align}
For pulses expressed in Cartesian coordinates, an additional hyperbolic tangent scaling function is applied to constrain the waveform amplitude within $[-1, 1]$:
\begin{align}
\label{scaleFunction}
x(t) = \frac{1 - e^{-\kappa x^s(t)}}{1 + e^{-\kappa x^s(t)}}.
\end{align}
Here, $\kappa$ controls the steepness of the transition period; a typical choice is $\kappa = 10$, which allows the pulse to reach its maximum amplitude. The gradient of the scaled waveform $\mathbf{x}$ with respect to the smoothed variables $\mathbf{x}^s$ is a diagonal matrix:
\begin{align}
\label{eq-scaleGradient}
\frac{d \mathbf{x}}{d \mathbf{x}^s} = \mathrm{diag}\left(\frac{2\kappa e^{-\kappa x^s_i}}{(1 + e^{-\kappa x^s_i})^2} \right), \quad i = 1, 2, \ldots, N.
\end{align}
By the chain rule, the gradient of the fidelity $\eta$ with respect to the control variables $\mathbf{x}^c$ is given by
\begin{align}
\label{totalGradient}
\frac{d \eta}{d \mathbf{x}^c} = \frac{d \eta}{d \mathbf{x}} \cdot \frac{d \mathbf{x}}{d \mathbf{x}^s} \cdot \frac{d \mathbf{x}^s}{d \mathbf{x}^c}.
\end{align}

\section{Numerical implementation}

\textcolor{RevisionColour}{The linear algebra was implemented in Liouville space using the irreducible spherical tensors as basis operators.} The global stiffness matrix was assembled using the function \texttt{sparse}, where the index matrices and value matrix were generated separately, and a three-dimensional array was employed to vectorize the computation of the value matrix. \textcolor{RevisionColour}{The element-wise multiplications in Eq.~\ref{eq-dKdxj} were vectorized using a diagonal block matrix. The details of matrix assembly and solving are provided in the supplementary material, Section S3.}

The control variables are updated using MMA, a widely used approach for large-scale, constrained nonlinear optimization problems, such as topology optimization \cite{sigmund2007Morphologybased, zhu2016Topology}. MMA is a gradient-based algorithm that leverages the values and gradients of the objective and constraint functions to iteratively construct and solve a sequence of convex subproblems. Each subproblem has a unique optimal solution that can be efficiently obtained via a dual approach \cite{svanberg1982Algorithma}. To employ the MMA algorithm, the maximization of the ensemble fidelity is formulated as a least-squares problem, i.e., by minimizing the following expression:
\begin{equation}
\begin{aligned}
\label{eq-lsqOriginal}
f(\mathbf{x}^c) = \sum_{k=1}^{N_{\text{ens}}} (1 - \eta_k)^2 
\end{aligned}.
\end{equation}
In the implementation, the least-squares objective was reformulated by converting the infidelities into $2 N_{\text{ens}}$ linear constraints while setting the objective function to zero\cite{svanberg2004some}:
\begin{equation}
\begin{aligned}
\label{eq-lsqFormulate}
& f_0(\mathbf{x}^c) = 0, \\
& f_k(\mathbf{x}^c) = 1 - \eta_k, \quad k = 1, 2, \ldots, N_{\text{ens}}, \\
& f_{N_{\text{ens}} + k}(\mathbf{x}^c) = \eta_k - 1, \quad k = 1, 2, \ldots, N_{\text{ens}}.
\end{aligned}
\end{equation}

\begin{algorithm}[!htb]
\caption{Find an optimal control pulse shape.}
\label{OptFlow}

\KwIn{Parameters of the spin system; Pulse parameters}
\KwOut{Optimized control pulse $\mathbf{x}$}

Initialize $\mathbf{x^c} \gets \texttt{rand}(M,N)$\tcp*[r]{Random initialization}
Solve $\eta_k$ from Eq.~\ref{eq-objective} and $d\eta_k/d\mathbf{x}$ from Eq.~\ref{eq-gradientByAdjoint}\;

\For{$iter \gets 2$ \textbf{to} $iter_{\text{max}}$}{
    \If{$\overline{\eta} \geq \text{target}$}{
        \textbf{break}\;
    }
    \Else{
        Compute $f_k$ from Eq.~\ref{eq-lsqFormulate} and $df_k/d\mathbf{x^c}$ from Eq.~\ref{totalGradient}\;
        Update $\mathbf{x^c}$ using MMA\tcp*[r]{Update control variables}   
        Solve $\mathbf{x^s}$ from Eq.~\ref{helmFEM} and $d\mathbf{x^s}/d\mathbf{x^c}$ from Eq.~\ref{helmGradient2}\tcp*[r]{Smooth variables} 
        \eIf{$\mathbf{x^c}$ are Cartesian components}{
            Solve $\mathbf{x}$ from Eq.~\ref{scaleFunction} and $d\mathbf{x}/d\mathbf{x^s}$ from Eq.~\ref{eq-scaleGradient} \tcp*[r]{Scale variables} 
        }{
            $\mathbf{x} \gets \mathbf{x^s}$\;
        }
        Solve $\bm{\alpha}$ from Algorithm~\ref{solveFEMrho}\tcp*[r]{Solve LvN equation} 
        Solve $\eta_k$ from Eq.~\ref{eq-objective} and $d\eta_k/d\mathbf{x}$ from Eq.~\ref{eq-gradientByAdjoint}\tcp*[r]{Adjoint analysis} 
    }
}
\Return{$\mathbf{x}$}\;
\end{algorithm}

Algorithm~\ref{OptFlow} outlines the pseudocode for solving the optimal control problem in Eq.~\ref{eq-optimalModel}. The control variables $\mathbf{x}^c$ consist of $M$ channels, each with $N$ discrete values. Iteration proceeds until the average fidelity $\overline{\eta}$ reaches the target (e.g., $0.995$) or the maximum number of iterations is reached. For an ensemble of spin systems, the mesh is kept identical across all members, so the index matrix of the stiffness matrix remains unchanged. Each ensemble member has an individual Liouvillian, and the corresponding stiffness matrix is assembled to evaluate the spin trajectory and its gradient. Handling an ensemble system can be efficiently accelerated using MATLAB’s parallel computing capabilities. In contrast, processing the variables involves solving a separate linear system, executed only once per iteration. For phase optimization, the smoothed variables $\mathbf{x}^s$ define the pulse shape; for Cartesian components ($x$ and $y$) optimization, $\mathbf{x}^s$ is further scaled to the range $[-1,1]$.

\section{Results and discussion}
\label{sec:performance}
\subsection{Accuracy and speed}
The accuracy and computational efficiency of the FEM approach for solving the single-spin system were evaluated. \textcolor{RevisionColour}{The step-by-step propagation method~\cite{khanejaOptimal2005} is adopted as a standard reference, where the control gradient is computed using the AUXMAT.} This referred method was executed using Spinach v2.8 \cite{hogben2011Spinach}, where step propagators are computed via the reordered Taylor expansion, summing low-order terms to machine precision ($2.22 \times 10^{-16}$ on a 64-bit machine).  \textcolor{RevisionColour}{All computations were performed in MATLAB 2025b on a PC equipped with an AMD Ryzen Threadripper PRO 3955WX 16-Cores processor (base frequency: 3.90 GHz) and 256 GB of RAM.} \textcolor{RevisionColour}{For benchmarking the speed of the FEM method, MATLAB's parallel pool was disabled to remove the influence of parallel computing.}

The relative error of the spin trajectory is defined as
\begin{align}
\label{eq-spinError}
\epsilon_{\rho} = \frac{1}{N+1} \sum_{i=1}^{N+1} \frac{|\rho_i^{\text{F}} - \rho_i^{\text{P}}|}{|\rho_i^{\text{P}}|},
\end{align}
where $N$ denotes the number of time steps, and $\rho_i^{\text{P}}$ and $\rho_i^{\text{F}}$ are the $i$-th spin vectors obtained from \textcolor{RevisionColour}{step-by-step propagation} and FEM, respectively. \textcolor{RevisionColour}{Since the spin state vectors are normalized, $|\rho_i^{\text{P}}|=1$.} And the relative error of the gradient is computed as
\begin{align}
\label{eq-gradError}
\epsilon_{\text{grad}} = \frac{1}{N} \sum_{i=1}^{N} \left| \frac{\nabla_i^{\text{F}} - \nabla_i^{\text{auxmat}}}{\nabla_i^{\text{auxmat}}} \right|,
\end{align}
where $\nabla_i^{\text{auxmat}}$ and $\nabla_i^{\text{F}}$ are the partial derivatives of the objective with respect to the $i$-th control variable, computed via \textcolor{RevisionColour}{AUXMAT} and FEM, respectively. The approximation error is mainly determined by the discrete step size $\|\mathbf{L}\| \Delta t$. In typical liquid-state NMR experiments, an RF amplitude of \SI{10}{kHz} and a time step of \SI{1}{\micro\second} yield $\|\mathbf{L}\| \Delta t = 0.063$. In this analysis, $\|\mathbf{L}\| \Delta t$ was varied from 0.01 to 0.1, and for each value, 30 random pulse shapes were generated to compute the mean values of $\epsilon_{\rho}$ and $\epsilon_{\text{grad}}$. 

\begin{figure}[htb]
    \includegraphics[width=\textwidth]{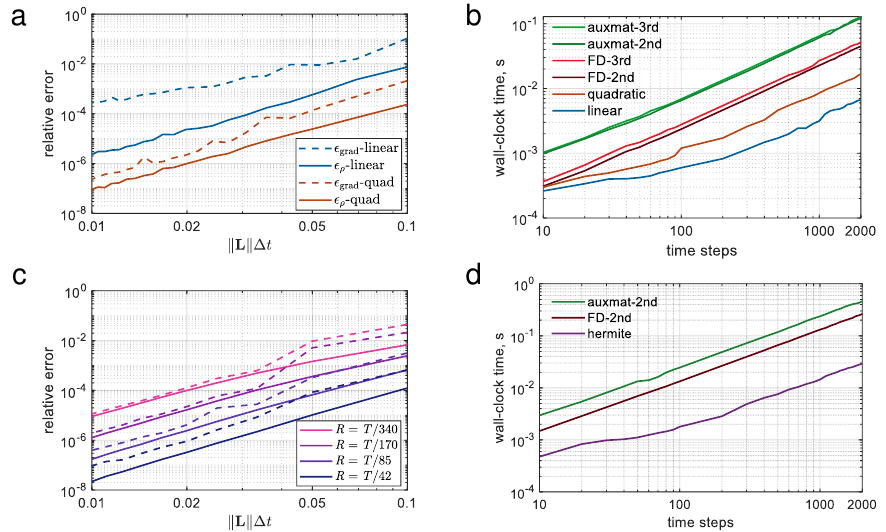}
    \caption{\textcolor{RevisionColour}{Performance of FEM and the step-by-step propagation method for computing the spin trajectory and gradient in a single-spin system.
 \textbf{Piecewise-constant waveform}: (a) Relative error for the FEM linear and quadratic elements. (b) Wall-clock time for the FEM linear and quadratic elements, AUXMAT, and 2nd-order finite difference (FD). The AUXMAT and FD methods used 2nd- and 3rd-order Taylor-truncated propagators, as indicated in the legend.  
 \textbf{Piecewise-linear waveform}: (c) Relative error for the FEM cubic Hermite elements, $R$ denotes the Helmholtz-filter radius, which controls waveform smoothing. Solid and dashed lines denote spin-trajectory and gradient error, respectively. (d) Wall-clock time for the FEM Hermite elements, AUXMAT, and FD methods. The AUXMAT and FD methods employed 2nd-order truncated propagators. Panels (a) and (c) use a pulse duration of $T=0.5$ ms, and a Liouvillian norm $\|\mathbf{L}\|=2\times10^{4}~\mathrm{rad}\cdot\mathrm{s}^{-1}$, with $N$ swept from 100 to 1000. Panels (b) and (d) fix $\|\mathbf{L}\|\Delta t=0.063$ and sweep $N$ from 10 to 2000.}}
    \label{fig-femErrorSpeed}
\end{figure}

\textcolor{RevisionColour}{Figure~\ref{fig-femErrorSpeed}a shows the benchmark results for linear and quadratic elements,} using the linear elements, the spin-trajectory error increases from $10^{-6}$ to $10^{-2}$, while the gradient error rises from $10^{-4}$ to $10^{-1}$. Employing quadratic shape functions significantly reduces these ranges to $10^{-7}$–$10^{-4}$ for the trajectory and $10^{-6}$–$10^{-3}$ for the gradient. An average ensemble fidelity of $0.995$ can be achieved with a trajectory error below $10^{-3}$, which for linear element approximations requires $||\mathbf{L}||\Delta t \leq 0.06$. \textcolor{RevisionColour}{The difference between $\epsilon_{\rho}$ and $\epsilon_{\text{grad}}$ arises because in Eq.~\ref{eq-spinError}, $|\rho_i^{\text{P}}|$ is fixed at 1, whereas in Eq.~\ref{eq-gradError}, $\nabla_i^{\text{auxmat}}$ can become very small for some control variables and reaches the approximation error of the linear or quadratic elements, resulting in an amplified relative error.} 

\textcolor{RevisionColour}{The performance of the linear and quadratic elements was compared with the AUXMAT and second-order finite difference (FD) methods, where the propagators were approximated by truncating the Taylor series. The linear elements achieve accuracy comparable to the AUXMAT/FD results obtained using second-order truncation, while the quadratic elements show comparable accuracy to AUXMAT/FD using third-order truncation (see supplementary material, Fig. S4). The corresponding wall-clock time as a function of time steps is presented in Fig.~\ref{fig-femErrorSpeed}(b). When $N$ ranges from 100 to 2000, the linear elements provide approximately a $7\times$ speedup over FD-2nd, and the quadratic elements yield roughly an $3.5\times$ speedup over the FD 3rd order approximation.}

\textcolor{RevisionColour}{The same procedure was applied to Hermite elements using a piecewise-linear waveform, and the relative error of the Hermite approximation is shown in Fig.~\ref{fig-femErrorSpeed}c. Unlike the case of piecewise-constant waveforms, the accuracy of the Hermite discretization depends on the smoothness of the linear waveform. Increasing the Helmholtz filter radius $R$ enforces smoother pulse shapes, reducing the error and exceeding the quadratic-element accuracy when $R = T/42$. And cubic Hermite elements offer an accuracy intermediate to AUXMAT/FD methods using second- and third-order Taylor truncation (see supplementary material, Fig. S5). The wall-clock time for evaluating gradients of the linear waveform is presented in Fig.~\ref{fig-femErrorSpeed}d, where the Hermite elements provide approximately a $10\times$ speedup over FD 2nd order approximation when $N$ ranges from 100 to 2000.} 

\textcolor{RevisionColour}{It is worth noting that the FEM wall-clock time is piecewise linear; this behavior arises because stiffness matrix assembly and linear solves are performed on large sparse matrices. As the number of time steps increases, the working set eventually exceeds the CPU cache capacity. The computation thus transitions from a compute-bound regime, which is dominated by small dense matrix operations, to a memory-bound regime limited by RAM bandwidth. Consequently, a few jumps appear in the runtime curve, and its slope increases slightly in the large-$N$ region.}

\subsection{State-to-state optimal control}
\begin{figure}[H]
    \centering\includegraphics[width=0.7\textwidth]{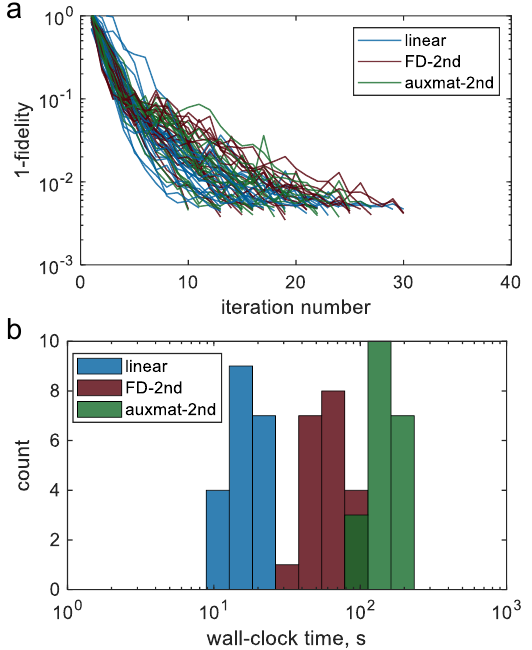}
    \caption{\textcolor{RevisionColour}{Performance of the excitation pulse optimization for an ensemble single-spin system.
    (a) Convergence rate comparison for gradients computed using FEM with linear elements, AUXMAT, and 2nd-order finite differences (FD). AUXMAT and FD used 2nd-order Taylor-truncated propagators. Each method was repeated 20 times from different random initial guesses, and the MMA algorithm was used for optimization.
    (b) Histogram showing the time consumption of the three methods. 
    The shaped pulse steers $\mathbf{I_{\text{z}}}$ to $\mathbf{I_{\text{x}}}$ with RF amplitude \SI{10}{kHz} and $\pm20\%$ scaling ($n_{\text{rf}}=5$). A \SI{20}{kHz} bandwidth was discretized into $n_{\text{off}}=51$ offsets, giving $N_{\text{ens}}=255$. The \SI{500}{\micro s} pulse was piecewise constant with 500 segments, its amplitudes were fixed at 1, and its phases were optimized. The target ensemble fidelity was 0.995. To eliminate the influence of parallel computation, the parallel pool was disabled in MATLAB 2025b.}}
    \label{fig-SS_FEMVSGRAPE}
\end{figure}

\textcolor{RevisionColour}{The FEM combined with MMA was tested by optimizing a single-spin broadband excitation pulse, which transfers $\mathbf{I_\mathrm{z}}$ to $\mathbf{I_\mathrm{x}}$ over a \SI{20}{kHz} offset range. The shaped pulse has a duration of \SI{500}{\micro s}, discretized into 500 time bins, and a nominal amplitude of \SI{10}{kHz} with $\pm 20\%$ variations. Gradients were computed using three methods: FEM with linear elements, the auxmat 2nd order, and FD 2nd order approximations, which yield comparable numerical accuracy. Each method was repeated 20 times from different random initial guesses, and their convergence rates are shown in Fig. \ref{fig-SS_FEMVSGRAPE}a. The similar convergence rates are primarily constrained by the MMA optimizer; its oscillatory convergence is analysed in the next section. In terms of runtime, the linear elements achieved an order-of-magnitude speedup over auxmat 2nd order approximation, while the FD 2nd order approximation exhibits intermediate wall-clock time, as shown in Fig. \ref{fig-SS_FEMVSGRAPE}b.} 

\textcolor{RevisionColour}{In the above test, a single-spin system was constructed for each ensemble member, and all fidelities and gradients obtained from either the FEM, AUXMAT, or FD methods were provided to the MMA algorithm, in which the ensemble constraints were built by Eq~\ref{eq-lsqFormulate}. When AUXMAT is combined with L-BFGS or Newton optimization, as implemented in Spinach, the Hamiltonians of the non-interacting subsystems can be assembled into a block-diagonal matrix, which allows the ensemble-averaged fidelity and gradient to be computed more efficiently. This block-diagonal acceleration was excluded from the comparisons.}

\subsection{Propagator optimization}
\begin{figure}[!htb]
    \centering
    \includegraphics[width=0.7\textwidth]{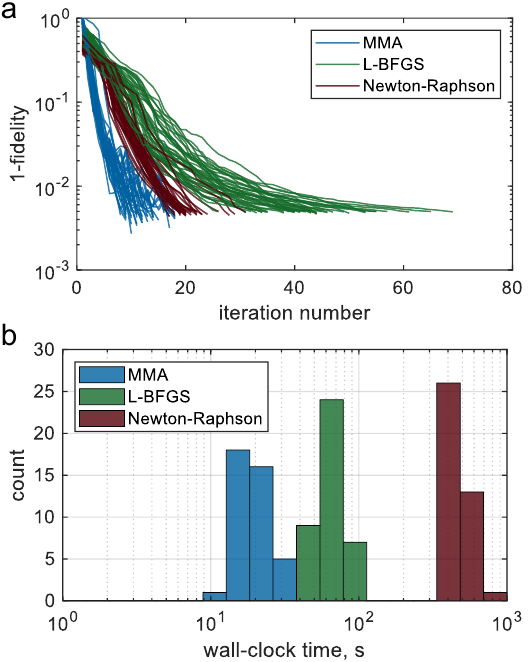}
    \caption{\textcolor{RevisionColour}{Performance of implementing a universal $90_\text{x}^\text{o}$ rotation for an ensemble single-spin system.
    (a) Comparison of convergence rates for the MMA, L-BFGS, and Newton-Raphson methods, each method was repeated 40 times using different random initial guesses, \textcolor{RevisionColour}{AUXMAT} was used to compute the gradient and Hessian.
    (b) Histogram of wall-clock time for three methods. The RF amplitude was \SI{10}{kHz} with $\pm10\%$ scaling ($n_\text{rf}=5$). A \SI{20}{kHz} bandwidth was sampled at $n_\text{off}=51$ offsets, giving an ensemble size $N_\text{ens}=255$. The \SI{500}{\micro s} pulse used 500 piecewise-constant slices with unit-fixed amplitude, and the phases were optimized. Optimization targeted an ensemble fidelity of 0.995 and terminated at 100 maximum iterations. Ensemble gradient and Hessian evaluations were accelerated using 16 workers in MATLAB R2025b.}}
    \label{fig-UR_mmaVSlbfgs}
\end{figure}

The performance of MMA was tested against the L-BFGS and Newton-Raphson methods, with the latter two executed using Spinach v2.8. The test case involved optimizing a broadband universal rotation pulse, designed to realize the target propagator $\mathbf{U} = \exp\left(-\text{i}\pi \mathbf{I_{\text{x}}}/2 \right)$ across a \SI{20}{kHz} offset range. The shaped pulse has a duration of \SI{500}{\micro s}, discretized into 500 time bins, with a nominal amplitude of \SI{10}{kHz} and $\pm 10\%$ scaling variations. The linear algebra was implemented in Hilbert space, where a unique target propagator favored by the optimization algorithm can be defined~\cite{kobzar2012Exploring}. Unlike the step-by-step propagation in the GRAPE method, the FEM model approximates spin evolution directly using a linear combination of basis functions. As a result, the effective propagator is not explicitly constructed, rendering propagator optimization currently infeasible within the FEM framework. \textcolor{RevisionColour}{AUXMAT was used to compute the gradient for both MMA and L-BFGS, and the Hessian matrix for the Newton-Raphson method; propagators were calculated using a Taylor series expansion truncated to machine precision.} The convergence behavior shown in Fig.~\ref{fig-UR_mmaVSlbfgs}a indicates that L-BFGS achieves stable but slower convergence as the fidelity approaches the target. The Newton method converges more rapidly and stably, while MMA achieves a comparable convergence rate to the Newton-Raphson method, albeit with a non-monotonic curve. The wall-clock times, shown in Fig.~\ref{fig-UR_mmaVSlbfgs}b, indicate that MMA completes the optimization faster than both L-BFGS and the Newton-Raphson methods. 

\textcolor{RevisionColour}{The oscillatory behavior of the MMA algorithm arises from the implicit Hessian matrix of its first-order approximation. The empirical choices of the moving asymptotes implicitly generate an approximated, positive-definite diagonal Hessian (curvature), which may be inaccurate. For example, loose asymptotes yield smaller curvature and consequently an
aggressive step that overshoots the local minimum. The algorithm then corrects this overshoot
by moving back in the next iteration, leading to an oscillatory pattern. The modified variant, globally convergent MMA (GCMMA)~\cite{svanbergClass2002}, addresses this by introducing an additional inner iteration that plays a role analogous to a line search. This process forces more conservative approximations of both the objective and constraints, yielding slower but more stable convergence. However, each inner iteration requires recomputing the ensemble fidelities, whose cost is comparable to computing both fidelities and gradients together.}

\section{Conclusion}
\label{sec:Conclusion}

In this work, we employed the FEM method to solve the Liouville–von Neumann equation for a single-spin system. \textcolor{RevisionColour}{Achieving gradient accuracy comparable to step-by-step propagation approach using second-order Taylor-truncated propagators, the use of linear elements provides an approximate order-of-magnitude speedup in gradient evaluation.} Additionally, the propagator optimization benchmark confirms that the MMA optimizer converges faster than L-BFGS, highlighting the potential of this approach for time-constrained optimal control problems.

\textcolor{RevisionColour}{With the adjoint method, gradient computation is as efficient as trajectory evaluation, which is similar to the case in step-by-step propagation. Hence, the computational cost is dominated by stiffness-matrix assembly and linear-system solving. While the 1-spin system attains the largest acceleration, this speedup reduces by approximately 4 for 2 spins (see supplementary material, Fig. S7) and is expected to diminish further with increasing spin system size, due to the runtime scaling with nearly the square of the nodal degrees of freedom in the current implementation. When the parallelization is used to accelerate the FEM and step-by-step propagation, it is also worth studying their relative efficiency as a function of the number of degrees of freedom.}

The oscillatory convergence behavior limits the applicability of MMA in high-fidelity scenarios, for example, $99.99\%$. A hybrid approach could use MMA to quickly reach an initial target and then switch to a more stable method for fine-tuning. Another direction is including spatial variables in the FEM model, either when diffusion becomes a degree of freedom~\cite{zientara1979Variational} or when optimal control of fluid samples~\cite{alinaghianjouzdaniOptimal2023} is concerned.

\section*{\textcolor{RevisionColour}{Supplementary Material}}
\textcolor{RevisionColour}{The supplementary material includes the derivation of stiffness matrices for quadratic and Hermite elements, and matrix assembly and solution in MATLAB. It also provides performance benchmarks for the FEM, auxiliary matrix formalism, and second-order finite difference methods: specifically evaluating accuracy at small step sizes and computational speed for 1- and 2-spin systems.}

\section*{Acknowledgements}
M.H.\ acknowledges support from the Joint Lab Virtual Materials Design (JLVMD) of the Helmholtz Association. J.G.K.\ acknowledges support from the ERC-SyG (HiSCORE, 951459). B.L.\ and J.G.K.\ acknowledge partial support from CRC 1527 HyPERiON. All authors acknowledge the Helmholtz Society's support through the Materials Systems Engineering program. Dr.\ Neil MacKinnon is sincerely thanked for discussions and editing the manuscript. 

\section*{Author contributions}
\textbf{Mengjia He:} Data curation (equal); Formal analysis (equal); Investigation (equal); Methodology (equal); Software (equal); Validation (equal); Visualization (equal); Writing - original draft (equal); Writing - review \& editing (equal). \textbf{Yongbo Deng:} Conceptualization (supporting); Formal analysis (equal); Investigation (supporting); Methodology (equal); Software (equal); Supervision (supporting); Validation (supporting); Writing - review \& editing (equal). \textbf{Burkhard Luy:} Conceptualization (supporting); Methodology (equal); Supervision (equal); Validation (supporting); Writing - review \& editing (equal). \textbf{Jan Korvink:} Conceptualization (equal); Formal analysis (equal); Funding acquisition (equal); Investigation (equal);
Methodology (equal); Project administration (equal); Resources (equal); Supervision (equal); Writing - review \& editing (equal).

\section*{Data availability}
\textcolor{RevisionColour}{The MATLAB code will be made available on GitHub at \url{https://github.com/kikioh}.}

\section*{Competing interests}
J.G.K. is a shareholder of Voxalytic GmbH. The other authors declare no competing interests. 

{\setstretch{0.9} 
\bibliography{ref.bib}
\bibliographystyle{unsrt}
}

\newpage
\section*{Supplementary Material}
\subsection*{S1. Stiffness matrix with quadratic elements}
\label{section-stiffmat-quad}
To improve accuracy, quadratic shape functions were employed. The element shape functions are given by
\begin{align}
\label{eq-quadShapeFunction}
H_1(\xi) = \frac{1}{2} \xi (\xi - 1), \quad
H_2(\xi) = 1 - \xi^2, \quad
H_3(\xi) = \frac{1}{2} \xi (\xi + 1), \quad \xi \in [-1,1].
\end{align}
Assuming the Liouvillian is constant within an element of size $2\Delta t$, the element stiffness matrix can be written as a $3\times3$ block matrix:
\begin{align}
\label{eq-quadEleStiff}
K^{e,j}_{mn} = \int_{-1}^1 \left[ \frac{dH_n}{d\xi}\frac{1}{\Delta t}+\text{i}\mathbf{L}_jH_n \right] H_m \cdot \Delta t d\xi,
\end{align} 
where $\mathbf{L}_j$ denotes the Liouvillian on the $j$th element and $m,n\in\{1,2,3\}$. The element stiffness matrix is therefore
\begin{align}
\label{eq-eleMatrix_quad}
K^{e,j} =
\begin{bmatrix}
-\frac{1}{2} &\frac{2}{3} &-\frac{1}{6}\\
-\frac{2}{3}&0&\frac{2}{3}\\
\frac{1}{6}&-\frac{2}{3}&\frac{1}{2} 
\end{bmatrix}\otimes \mathbf{E}+
\begin{bmatrix}
\frac{4}{15}&\frac{2}{15} &-\frac{1}{15}\\
\frac{2}{15} &\frac{16}{15}&\frac{2}{15}\\
-\frac{1}{15}&\frac{2}{15} &\frac{4}{15}
\end{bmatrix} \otimes \mathrm{i}\mathbf{L}\Delta t
\end{align}
where $\mathbf{E}$ denotes the identity matrix. The global stiffness matrix is given by
\begin{align}
\label{rhoStiff_quad}
\mathbf{K} =
\begin{bmatrix}
K_{11}^{e,1} & K_{12}^{e,1} & K_{13}^{e,1} &  &  &  &  \\
K_{21}^{e,1} & K_{22}^{e,1} & K_{23}^{e,1} &  &  &  & \\
K_{31}^{e,1} & K_{32}^{e,1} & K_{33}^{e,1} + K_{11}^{e,2} & K_{12}^{e,2} & &  \\
 &  & K_{21}^{e,2} & \ddots  & \ddots &  \\
&  &  & \ddots & K_{33}^{e,N-1}+K_{11}^{e,N} & K_{12}^{e,N} & K_{13}^{e,N} \\
 &  &  &  & K_{21}^{e,N}  & K_{22}^{e,N}& K_{23}^{e,N} \\
 &  &  &  & K_{31}^{e,N}  & K_{32}^{e,N}& K_{33}^{e,N}
\end{bmatrix}.
\end{align}
For a single-spin system, the Liouvillian $\mathbf{L}\in\mathbb{C}^{4\times 4}$, element stiffness matrix $K^{e,j}\in\mathbb{C}^{12\times 12}$ connects three nodes, and the global stiffness matrix $\mathbf{K}\in\mathbb{C}^{(8N+4)\times(8N+4)}$.

\newpage
\subsection*{S2. Stiffness matrix with Hermite elements}
\label{section-stiffmat-hermite}
\begin{figure}[htbp]
\renewcommand{\thefigure}{S1}
    \centering
    \includegraphics[width=0.9\textwidth]{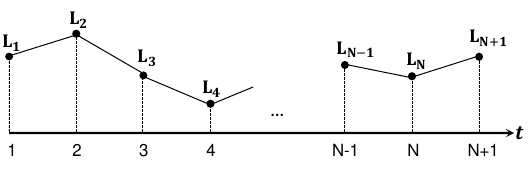}
    \caption{View of the piecewise-linear waveform, the  Liouvillians are defined on the nodes.}
    \label{fig-pw-linear}
\end{figure}
If one needs to optimize a piecewise-linear waveform in which the Liouvillian is continuous across time intervals, which is shown in Fig.~\ref{fig-pw-linear}. The cubic Hermite shape functions enforce $C^1$ continuity between elements and are therefore appropriate for this case. The element shape functions are given by
\begin{align}
\label{eq-hermiteShapeFunction}
H_1(\xi) &= \frac{1}{4}(1-\xi)^2 (2+\xi), \quad
H_2(\xi) = \frac{1}{4}(1-\xi)^2 (1+\xi), \notag \\[4pt]
H_3(\xi) &= \frac{1}{4}(1+\xi)^2 (2-\xi), \quad
H_4(\xi) = \frac{1}{4}(1+\xi)^2 (\xi-1), \quad \xi \in [-1,1].
\end{align}
The element stiffness matrix can be expressed as a $4 \times 4$ block matrix, i.e.,
\begin{align}
\label{eq-hermiteEleStiff}
K^{e,j}_{mn} = \int_{-1}^1 \left[ \frac{dH_n}{d\xi}\frac{2}{\Delta t}+\text{i}\left(\mathbf{L}_j+\frac{\xi+1}{2}(\mathbf{L}_{j+1}-\mathbf{L}_j)\right) H_n \right] H_m \cdot \frac{\Delta t}{2} d\xi,
\end{align} 
where $\mathbf{L}_j$ and $\mathbf{L}_{j+1}$ denote the Liouvillians at the left and right nodes of the $j$th element, respectively, and $m,n\in\{1,2,3,4\}$. The element stiffness matrix is computed as
\begin{align}
\label{eq-eleMartix_hermite}
K^{e,j}=\scalebox{0.7}{$\begin{bmatrix}
-\frac{1}{2} & \frac{1}{5} & \frac{1}{2} & -\frac{1}{5} \\
-\frac{1}{5} & 0 & \frac{1}{5} & -\frac{1}{15} \\
-\frac{1}{2} & -\frac{1}{5} & \frac{1}{2} & \frac{1}{5} \\
\frac{1}{5} & \frac{1}{15} & -\frac{1}{5} & 0
\end{bmatrix}$}  \otimes \mathbf{E}  + 
\scalebox{0.7}{$\begin{bmatrix}
\frac{2}{7} & \frac{1}{14} & \frac{9}{140} & -\frac{1}{30} \\
\frac{1}{14} & \frac{1}{42} & \frac{1}{35} & -\frac{1}{70} \\
\frac{9}{140} & \frac{1}{35} & \frac{3}{35} & -\frac{1}{30} \\
-\frac{1}{30} & -\frac{1}{70} & -\frac{1}{30} & \frac{1}{70}
\end{bmatrix}$} \otimes \mathrm{i}\Delta t \mathbf{L}_j
+
\scalebox{0.7}{$\begin{bmatrix}
\frac{3}{35} & \frac{1}{30} & \frac{9}{140} & -\frac{1}{35} \\
\frac{1}{30} & \frac{1}{70} & \frac{1}{30} & -\frac{1}{70} \\
\frac{9}{140} & \frac{1}{30} & \frac{2}{7} & -\frac{1}{14} \\
-\frac{1}{35} & -\frac{1}{70} & -\frac{1}{14} & \frac{1}{42}
\end{bmatrix}$} \otimes \mathrm{i}\Delta t \mathbf{L}_{j+1}.
\end{align}
And the global stiffness matrix is given by
\begin{align}
\label{rhoStiff_hermite}
\mathbf{K} =\scalebox{0.8}{$
\begin{bmatrix}
K_{11}^{e,1} & K_{12}^{e,1} & K_{13}^{e,1} & K_{14}^{e,1} & & & & & \\
K_{21}^{e,1} & K_{22}^{e,1} & K_{23}^{e,1} & K_{24}^{e,1} & & & & & \\
K_{31}^{e,1} & K_{32}^{e,1} & K_{33}^{e,1} + K_{11}^{e,2} & K_{34}^{e,1} + K_{12}^{e,2} & K_{13}^{e,2} & K_{14}^{e,2} & & & \\
K_{41}^{e,1} & K_{42}^{e,1} & K_{43}^{e,1} + K_{21}^{e,2} & K_{44}^{e,1} + K_{22}^{e,2} & K_{23}^{e,2} & K_{24}^{e,2} & & & \\
& & K_{31}^{e,2} & K_{32}^{e,2} & \ddots & & & & \\
& & K_{41}^{e,2} & K_{42}^{e,2} & & \ddots & & & \\
& & & & & \ddots & \ddots & \ddots & & \\
& & & & & & \ddots & \ddots & \ddots & \\
& & & & & & K_{33}^{e,N-1}+K_{11}^{e,N} & K_{34}^{e,N-1}+ K_{12}^{e,N} & K_{13}^{e,N} & K_{14}^{e,N}\\
& & & & & & K_{43}^{e,N-1}+K_{21}^{e,N} & K_{44}^{e,N-1}+K_{22}^{e,N} & K_{23}^{e,N} & K_{24}^{e,N} \\
& & & & & & K_{31}^{e,N} & K_{32}^{e,N} & K_{33}^{e,N} & K_{34}^{e,N}\\
& & & & & & K_{41}^{e,N} & K_{42}^{e,N} & K_{43}^{e,N} & K_{44}^{e,N}
\end{bmatrix}$}.
\end{align}
For a single-spin system, the Liouvillian $\mathbf{L}\in\mathbb{C}^{4\times 4}$, the element stiffness matrix $K^{e,j}\in\mathbb{C}^{16\times 16}$, and the global stiffness matrix $\mathbf{K}\in\mathbb{C}^{(8N+8)\times(8N+8)}$.

\subsection*{S3. Matrix assembly and solving}
\label{section-matrixSolving}
\begin{figure}[!htbp]
\renewcommand{\thefigure}{S2}
\centering
\includegraphics[width=\textwidth]{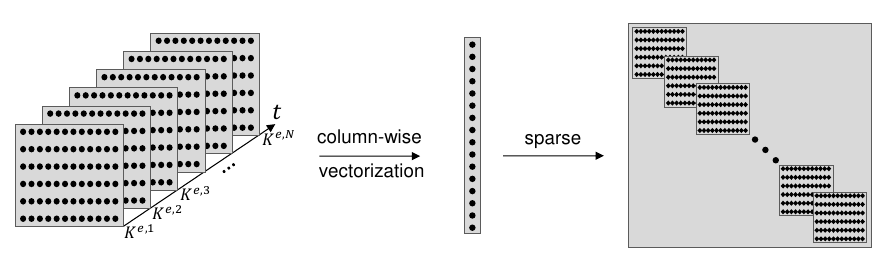}
\caption{Assembly of the stiffness matrix in MATLAB. The element stiffness matrices were stored as a 3D array, vectorized, and then transformed into a 2D sparse matrix using the \texttt{sparse} function.}
\label{fig-Kmatrix_assembly}
\end{figure}

Depending on the interpolation element type, the equation \ref{eq-eleMatrix_quad}, or \ref{eq-eleMartix_hermite} was used to generate the element stiffness matrices from the piecewise Liouvillians. As illustrated in Fig.~\ref{fig-Kmatrix_assembly}, these element matrices were computed in a vectorized manner and stored in a three-dimensional array, where each slice corresponds to one time segment. This 3D array is then vectorized and assembled into a 2D sparse matrix using MATLAB’s \texttt{sparse} function, with the index vectors generated separately.

The spin trajectory and the adjoint vector are jointly solved using an LU decomposition, taking advantage of the banded sparse matrix structure:
\begin{center}
\begin{minipage}{0.5\textwidth}
\begin{verbatim}
dK = decomposition(K, 'banded');
alpha = dK \ f;
lambda = eta / dK;
\end{verbatim}
\end{minipage}
\end{center}
Here, \texttt{eta} denotes the extended objective.

\begin{figure}[htbp]
\centering
\includegraphics[width=0.6\textwidth]{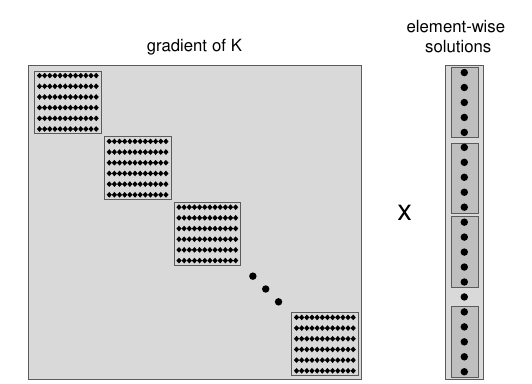}
\renewcommand{\thefigure}{S3}
\caption{Computation of the gradient of the stiffness matrix with respect to the control variables. The left-hand gradient of $\mathbf{K}$ is constructed by stacking the analytic gradient matrices of $K^e$ into a block-diagonal form, while the right-hand vector represents the element-wise spin states corresponding to each $K^e$ on the left.}
\label{fig-KG_assembly}
\end{figure}

The gradient of the stiffness matrix is evaluated by multiplying a block-diagonal matrix with a vector containing the element-wise solutions, as illustrated in Fig.~\ref{fig-KG_assembly}. Each block in the matrix represents the analytical gradient of a piecewise Liouvillian with respect to the local control parameters. The right-hand vector contains the state solutions obtained from the spin trajectory solution, and the corresponding degrees of freedom are arranged to align with the left-hand matrix according to the gradient of the element stiffness matrix.

\subsection*{S4. Accuracy of FEM, auxiliary matrix formalism, and second-order finite difference}

\begin{figure}[H]
\renewcommand{\thefigure}{S4}
\centering
\includegraphics[width=\textwidth]{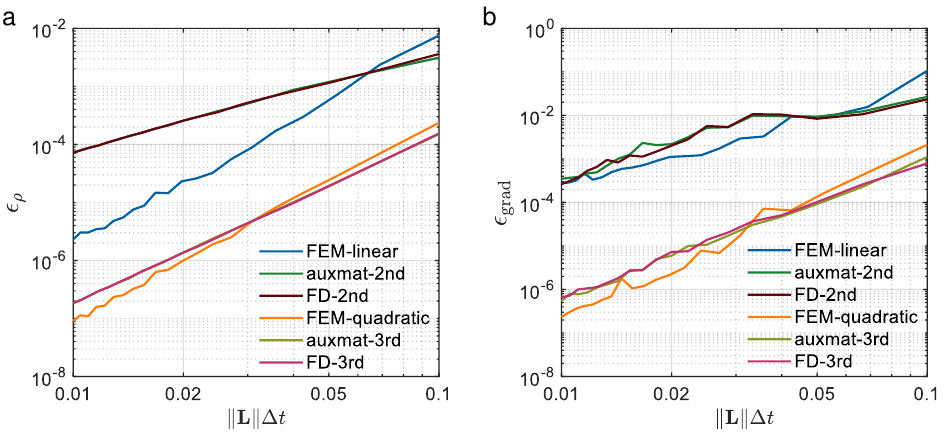}
\caption{The relative spin trajectory (a) and gradient (b) errors for the three compared methods. These errors are calculated relative to a reference solution obtained using machine-precision step-by-step propagation combined with the auxiliary matrix formalism (AUXMAT) for gradient computation. The computation considered a single-spin system driven by a piecewise-constant waveform. In approximated AUXMAT and the second-order central finite difference (FD) method, the propagators were evaluated via Taylor-series expansions truncated at second and third order, as indicated in the legends. Note that with second or third order Taylor truncation, the errors of AUXMAT and FD nearly coincide because they are dominated by the truncation error in propagator computation. The finite difference used a differentiation step of $h=0.1$. The pulse duration was set to $T=0.5$ ms, the Liouvillian norm was $\|\mathbf{L}\|=2\times10^{4}~\mathrm{rad}\cdot\mathrm{s}^{-1}$, and the number of time slices $N$ was swept from 100 to 1000.}
\label{fig-error_fem_auxmat_FD_rect}
\end{figure}

\begin{figure}[H]
\renewcommand{\thefigure}{S5}
\centering
\includegraphics[width=\textwidth]{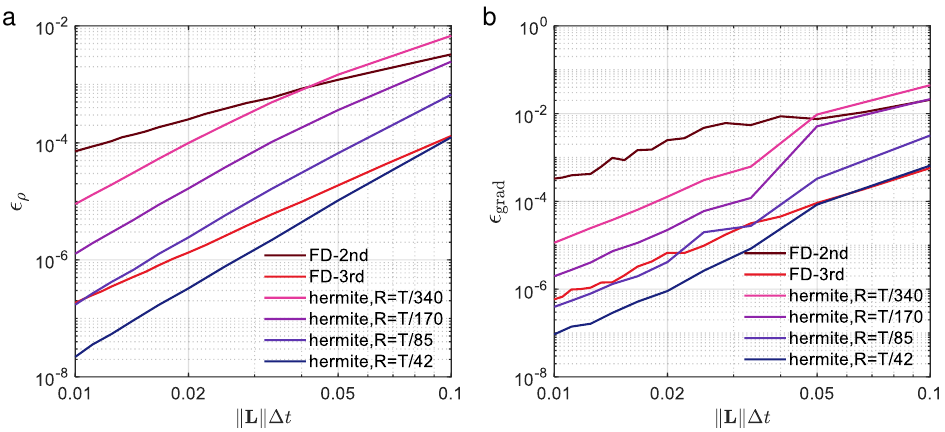}
\caption{The relative errors of the spin trajectory (a) and gradient (b) for three methods are shown, using a piecewise-linear waveform while keeping all other parameters identical to Fig.~\ref{fig-error_fem_auxmat_FD_rect}. Results from auxmat-2nd and auxmat-3rd are not shown because their errors are nearly identical to FD-2nd and FD-3rd, respectively. Note that the errors of the cubic Hermite elements depend on the pulse smoothness, where $T$ is the pulse duration and $R$ is the Helmholtz-filter radius controlling the smoothing.}
\label{fig-error_fem_auxmat_FD_trap}
\end{figure}
\clearpage

\subsection*{S5. Speed of FEM, auxiliary matrix formalism, and second-order finite difference}

\begin{figure}[H]
\renewcommand{\thefigure}{S6}
\centering
\includegraphics[width=\textwidth]{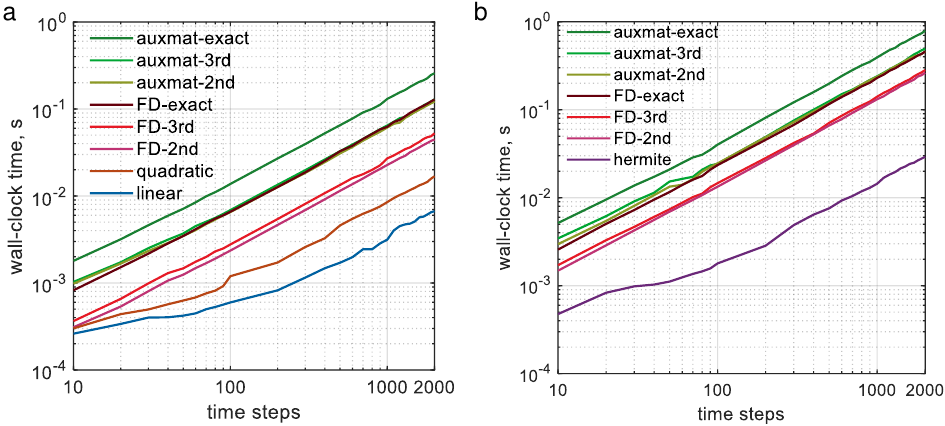}
\caption{Wall-clock time for computing spin trajectories and derivatives in a single-spin system. (a) Running time for the piecewise-constant waveform. (b) Running time for the piecewise-linear waveform. In both panels, $\|\mathbf{L}\|\Delta t = 0.063$ is held fixed, and the number of time steps $N$ is swept from 10 to 2000. MATLAB’s parallel pool was disabled to eliminate the influence of parallel execution. Each benchmark was repeated 30 times for each method, and the average running time was recorded. Across the practically relevant regime of $N = 100$ to $N = 2000$, linear elements achieve an approximate $7\times$ acceleration relative to FD-2nd; quadratic elements achieve $3.5\times$ acceleration relative to FD-3rd; and cubic Hermite elements achieve about $10\times$ acceleration relative to FD-3rd.}
\label{fig-speed_fem_auxmat_FD_1spin}
\end{figure}

\begin{figure}[H]
\renewcommand{\thefigure}{S7}
\centering
\includegraphics[width=\textwidth]{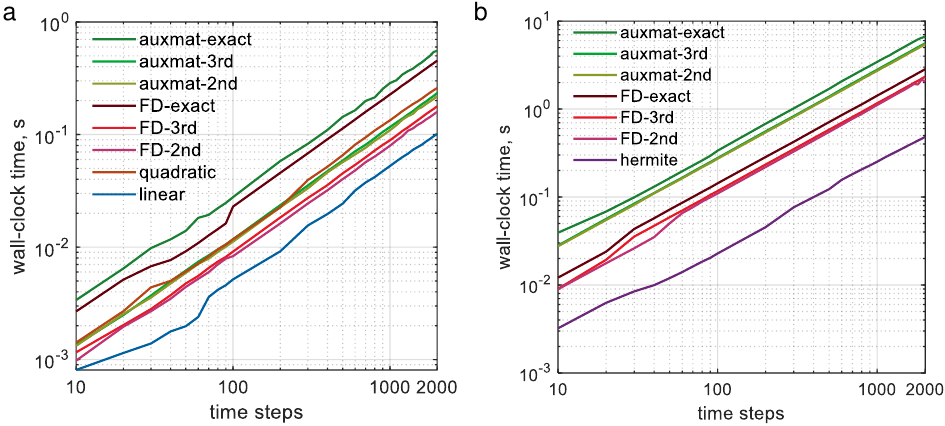}
\caption{Wall-clock time benchmarks for computing spin trajectories and gradients in a 2-spin system. (a) Running times for a piecewise-constant waveform. (b) Running times for a piecewise-linear waveform. Within the range of $100 \leq N \leq 2000$, linear elements show a $1.6\times$ reduction in running time relative to the FD-2nd, quadratic elements give comparable running times to FD-3rd, and cubic Hermite elements give an approximate $5\times$ speedup relative to FD-3rd.}
\label{fig-speed_fem_auxmat_FD_2spin}
\end{figure}
\end{document}